\documentclass[conference,10pt,twoside,twocolumn]{IEEEtran}

\usepackage{cite}
\usepackage[T1]{fontenc}
\usepackage{graphicx}
\usepackage{amssymb}
\usepackage{amsmath}
\usepackage{amsthm}
\usepackage{subfigure}
\usepackage{booktabs} 
\usepackage{multirow}
\usepackage{microtype}
\usepackage{balance}
\usepackage{xcolor}

\usepackage{amssymb}
\usepackage{amsfonts}
\usepackage{mathrsfs}
\usepackage{xspace}
\usepackage{bm}
\usepackage{upgreek}

\newcommand{\safemath}[2]{\newcommand{#1}{\ensuremath{#2}\xspace}}

\safemath{\bma}{\mathbf{a}}
\safemath{\bmb}{\mathbf{b}}
\safemath{\bmc}{\mathbf{c}}
\safemath{\bmd}{\mathbf{d}}
\safemath{\bme}{\mathbf{e}}
\safemath{\bmf}{\mathbf{f}}
\safemath{\bmg}{\mathbf{g}}
\safemath{\bmh}{\mathbf{h}}
\safemath{\bmi}{\mathbf{i}}
\safemath{\bmj}{\mathbf{j}}
\safemath{\bmk}{\mathbf{k}}
\safemath{\bml}{\mathbf{l}}
\safemath{\bmm}{\mathbf{m}}
\safemath{\bmn}{\mathbf{n}}
\safemath{\bmo}{\mathbf{o}}
\safemath{\bmp}{\mathbf{p}}
\safemath{\bmq}{\mathbf{q}}
\safemath{\bmr}{\mathbf{r}}
\safemath{\bms}{\mathbf{s}}
\safemath{\bmt}{\mathbf{t}}
\safemath{\bmu}{\mathbf{u}}
\safemath{\bmv}{\mathbf{v}}
\safemath{\bmw}{\mathbf{w}}
\safemath{\bmx}{\mathbf{x}}
\safemath{\bmy}{\mathbf{y}}
\safemath{\bmz}{\mathbf{z}}
\safemath{\bmzero}{\mathbf{0}}
\safemath{\bmone}{\mathbf{1}}

\bmdefine{\biad}{a}
\bmdefine{\bibd}{b}
\bmdefine{\bicd}{c}
\bmdefine{\bidd}{d}
\bmdefine{\bied}{e}
\bmdefine{\bifd}{f}
\bmdefine{\bigd}{g}
\bmdefine{\bihd}{h}
\bmdefine{\biid}{i}
\bmdefine{\bijd}{j}
\bmdefine{\bikd}{k}
\bmdefine{\bild}{l}
\bmdefine{\bimd}{m}
\bmdefine{\bind}{n}
\bmdefine{\biod}{o}
\bmdefine{\bipd}{p}
\bmdefine{\biqd}{q}
\bmdefine{\bird}{r}
\bmdefine{\bisd}{s}
\bmdefine{\bitd}{t}
\bmdefine{\biud}{u}
\bmdefine{\bivd}{v}
\bmdefine{\biwd}{w}
\bmdefine{\bixd}{x}
\bmdefine{\biyd}{y}
\bmdefine{\bizd}{z}

\bmdefine{\bixid}{\xi}
\bmdefine{\bilambdad}{\lambda}
\bmdefine{\bimud}{\mu}
\bmdefine{\bithetad}{\theta}
\bmdefine{\biphid}{\phi}
\bmdefine{\bideltad}{\delta}

\safemath{\bmia}{\biad}
\safemath{\bmib}{\bibd}
\safemath{\bmic}{\bicd}
\safemath{\bmid}{\bidd}
\safemath{\bmie}{\bied}
\safemath{\bmif}{\bifd}
\safemath{\bmig}{\bigd}
\safemath{\bmih}{\bihd}
\safemath{\bmii}{\biid}
\safemath{\bmij}{\bijd}
\safemath{\bmik}{\bikd}
\safemath{\bmil}{\bild}
\safemath{\bmim}{\bimd}
\safemath{\bmin}{\bind}
\safemath{\bmio}{\biod}
\safemath{\bmip}{\bipd}
\safemath{\bmiq}{\biqd}
\safemath{\bmir}{\bird}
\safemath{\bmis}{\bisd}
\safemath{\bmit}{\bitd}
\safemath{\bmiu}{\biud}
\safemath{\bmiv}{\bivd}
\safemath{\bmiw}{\biwd}
\safemath{\bmix}{\bixd}
\safemath{\bmiy}{\biyd}
\safemath{\bmiz}{\bizd}

\safemath{\bmxi}{\bixid}
\safemath{\bmlambda}{\bilambdad}
\safemath{\bmmu}{\bimud}
\safemath{\bmtheta}{\bithetad}
\safemath{\bmphi}{\biphid}
\safemath{\bmdelta}{\bideltad}

\safemath{\bA}{\mathbf{A}}
\safemath{\bB}{\mathbf{B}}
\safemath{\bC}{\mathbf{C}}
\safemath{\bD}{\mathbf{D}}
\safemath{\bE}{\mathbf{E}}
\safemath{\bF}{\mathbf{F}}
\safemath{\bG}{\mathbf{G}}
\safemath{\bH}{\mathbf{H}}
\safemath{\bI}{\mathbf{I}}
\safemath{\bJ}{\mathbf{J}}
\safemath{\bK}{\mathbf{K}}
\safemath{\bL}{\mathbf{L}}
\safemath{\bM}{\mathbf{M}}
\safemath{\bN}{\mathbf{N}}
\safemath{\bO}{\mathbf{O}}
\safemath{\bP}{\mathbf{P}}
\safemath{\bQ}{\mathbf{Q}}
\safemath{\bR}{\mathbf{R}}
\safemath{\bS}{\mathbf{S}}
\safemath{\bT}{\mathbf{T}}
\safemath{\bU}{\mathbf{U}}
\safemath{\bV}{\mathbf{V}}
\safemath{\bW}{\mathbf{W}}
\safemath{\bX}{\mathbf{X}}
\safemath{\bY}{\mathbf{Y}}
\safemath{\bZ}{\mathbf{Z}}

\safemath{\bZero}{\mathbf{0}}
\safemath{\bOne}{\mathbf{1}}
\safemath{\bDelta}{\mathbf{\Delta}}
\safemath{\bLambda}{\mathbf{\UpLambda}}
\safemath{\bPhi}{\mathbf{\Upphi}}
\safemath{\bSigma}{\mathbf{\Upsigma}}
\safemath{\bOmega}{\mathbf{\Upomega}}
\safemath{\bTheta}{\mathbf{\Uptheta}}

\bmdefine{\biAd}{A}
\bmdefine{\biBd}{B}
\bmdefine{\biCd}{C}
\bmdefine{\biDd}{D}
\bmdefine{\biEd}{E}
\bmdefine{\biFd}{F}
\bmdefine{\biGd}{G}
\bmdefine{\biHd}{H}
\bmdefine{\biId}{I}
\bmdefine{\biJd}{J}
\bmdefine{\biKd}{K}
\bmdefine{\biLd}{L}
\bmdefine{\biMd}{M}
\bmdefine{\biOd}{N}
\bmdefine{\biPd}{O}
\bmdefine{\biQd}{P}
\bmdefine{\biRd}{R}
\bmdefine{\biSd}{S}
\bmdefine{\biTd}{T}
\bmdefine{\biUd}{U}
\bmdefine{\biVd}{V}
\bmdefine{\biWd}{W}
\bmdefine{\biXd}{X}
\bmdefine{\biYd}{Y}
\bmdefine{\biZd}{Z}

\bmdefine{\biDelta}{\Delta}
\bmdefine{\biLambda}{\Lambda}
\bmdefine{\biPhi}{\Phi}
\bmdefine{\biSigma}{\Sigma}
\bmdefine{\biOmega}{\Omega}
\bmdefine{\biTheta}{\Theta}

\safemath{\bimA}{\biAd}
\safemath{\bimB}{\biBd}
\safemath{\bimC}{\biCd}
\safemath{\bimD}{\biDd}
\safemath{\bimE}{\biEd}
\safemath{\bimF}{\biFd}
\safemath{\bimG}{\biGd}
\safemath{\bimH}{\biHd}
\safemath{\bimI}{\biId}
\safemath{\bimJ}{\biJd}
\safemath{\bimK}{\biKd}
\safemath{\bimL}{\biLd}
\safemath{\bimM}{\biMd}
\safemath{\bimN}{\biNd}
\safemath{\bimO}{\biOd}
\safemath{\bimP}{\biPd}
\safemath{\bimQ}{\biQd}
\safemath{\bimR}{\biRd}
\safemath{\bimS}{\biSd}
\safemath{\bimT}{\biTd}
\safemath{\bimU}{\biUd}
\safemath{\bimV}{\biVd}
\safemath{\bimW}{\biWd}
\safemath{\bimX}{\biXd}
\safemath{\bimY}{\biYd}
\safemath{\bimZ}{\biZd}

\safemath{\bimDelta}{\biDelta}
\safemath{\bimLambda}{\biLambda}
\safemath{\bimPhi}{\biPhi}
\safemath{\bimSigma}{\biSigma}
\safemath{\bimOmega}{\biOmega}
\safemath{\bimTheta}{\biTheta}

\safemath{\setA}{\mathcal{A}}
\safemath{\setB}{\mathcal{B}}
\safemath{\setC}{\mathcal{C}}
\safemath{\setD}{\mathcal{D}}
\safemath{\setE}{\mathcal{E}}
\safemath{\setF}{\mathcal{F}}
\safemath{\setG}{\mathcal{G}}
\safemath{\setH}{\mathcal{H}}
\safemath{\setI}{\mathcal{I}}
\safemath{\setJ}{\mathcal{J}}
\safemath{\setK}{\mathcal{K}}
\safemath{\setL}{\mathcal{L}}
\safemath{\setM}{\mathcal{M}}
\safemath{\setN}{\mathcal{N}}
\safemath{\setO}{\mathcal{O}}
\safemath{\setP}{\mathcal{P}}
\safemath{\setQ}{\mathcal{Q}}
\safemath{\setR}{\mathcal{R}}
\safemath{\setS}{\mathcal{S}}
\safemath{\setT}{\mathcal{T}}
\safemath{\setU}{\mathcal{U}}
\safemath{\setV}{\mathcal{V}}
\safemath{\setW}{\mathcal{W}}
\safemath{\setX}{\mathcal{X}}
\safemath{\setY}{\mathcal{Y}}
\safemath{\setZ}{\mathcal{Z}}
\safemath{\emptySet}{\varnothing}

\safemath{\colA}{\mathscr{A}}
\safemath{\colB}{\mathscr{B}}
\safemath{\colC}{\mathscr{C}}
\safemath{\colD}{\mathscr{D}}
\safemath{\colE}{\mathscr{E}}
\safemath{\colF}{\mathscr{F}}
\safemath{\colG}{\mathscr{G}}
\safemath{\colH}{\mathscr{H}}
\safemath{\colI}{\mathscr{I}}
\safemath{\colJ}{\mathscr{J}}
\safemath{\colK}{\mathscr{K}}
\safemath{\colL}{\mathscr{L}}
\safemath{\colM}{\mathscr{M}}
\safemath{\colN}{\mathscr{N}}
\safemath{\colO}{\mathscr{O}}
\safemath{\colP}{\mathscr{P}}
\safemath{\colQ}{\mathscr{Q}}
\safemath{\colR}{\mathscr{R}}
\safemath{\colS}{\mathscr{S}}
\safemath{\colT}{\mathscr{T}}
\safemath{\colU}{\mathscr{U}}
\safemath{\colV}{\mathscr{V}}
\safemath{\colW}{\mathscr{W}}
\safemath{\colX}{\mathscr{X}}
\safemath{\colY}{\mathscr{Y}}
\safemath{\colZ}{\mathscr{Z}}

\safemath{\opA}{\mathbb{A}}
\safemath{\opB}{\mathbb{B}}
\safemath{\opC}{\mathbb{C}}
\safemath{\opD}{\mathbb{D}}
\safemath{\opE}{\mathbb{E}}
\safemath{\opF}{\mathbb{F}}
\safemath{\opG}{\mathbb{G}}
\safemath{\opH}{\mathbb{H}}
\safemath{\opI}{\mathbb{I}}
\safemath{\opJ}{\mathbb{J}}
\safemath{\opK}{\mathbb{K}}
\safemath{\opL}{\mathbb{L}}
\safemath{\opM}{\mathbb{M}}
\safemath{\opN}{\mathbb{N}}
\safemath{\opO}{\mathbb{O}}
\safemath{\opP}{\mathbb{P}}
\safemath{\opQ}{\mathbb{Q}}
\safemath{\opR}{\mathbb{R}}
\safemath{\opS}{\mathbb{S}}
\safemath{\opT}{\mathbb{T}}
\safemath{\opU}{\mathbb{U}}
\safemath{\opV}{\mathbb{V}}
\safemath{\opW}{\mathbb{W}}
\safemath{\opX}{\mathbb{X}}
\safemath{\opY}{\mathbb{Y}}
\safemath{\opZ}{\mathbb{Z}}
\safemath{\opZero}{\mathbb{O}}
\safemath{\identityop}{\opI}

\safemath{\veca}{\bma}
\safemath{\vecb}{\bmb}
\safemath{\vecc}{\bmc}
\safemath{\vecd}{\bmd}
\safemath{\vece}{\bme}
\safemath{\vecf}{\bmf}
\safemath{\vecg}{\bmg}
\safemath{\vech}{\bmh}
\safemath{\veci}{\bmi}
\safemath{\vecj}{\bmj}
\safemath{\veck}{\bmk}
\safemath{\vecl}{\bml}
\safemath{\vecm}{\bmm}
\safemath{\vecn}{\bmn}
\safemath{\veco}{\bmo}
\safemath{\vecp}{\bmp}
\safemath{\vecq}{\bmq}
\safemath{\vecr}{\bmr}
\safemath{\vecs}{\bms}
\safemath{\vect}{\bmt}
\safemath{\vecu}{\bmu}
\safemath{\vecv}{\bmv}
\safemath{\vecw}{\bmw}
\safemath{\vecx}{\bmx}
\safemath{\vecy}{\bmy}
\safemath{\vecz}{\bmz}

\safemath{\veczero}{\bmzero}
\safemath{\vecone}{\bmone}
\safemath{\vecxi}{\bmxi}
\safemath{\veclambda}{\bmlambda}
\safemath{\vecmu}{\bmmu}
\safemath{\vectheta}{\bmtheta}
\safemath{\vecphi}{\bmphi}
\safemath{\vecdelta}{\bmdelta}

\safemath{\matA}{\bA}
\safemath{\matB}{\bB}
\safemath{\matC}{\bC}
\safemath{\matD}{\bD}
\safemath{\matE}{\bE}
\safemath{\matF}{\bF}
\safemath{\matG}{\bG}
\safemath{\matH}{\bH}
\safemath{\matI}{\bI}
\safemath{\matJ}{\bJ}
\safemath{\matK}{\bK}
\safemath{\matL}{\bL}
\safemath{\matM}{\bM}
\safemath{\matN}{\bN}
\safemath{\matO}{\bO}
\safemath{\matP}{\bP}
\safemath{\matQ}{\bQ}
\safemath{\matR}{\bR}
\safemath{\matS}{\bS}
\safemath{\matT}{\bT}
\safemath{\matU}{\bU}
\safemath{\matV}{\bV}
\safemath{\matW}{\bW}
\safemath{\matX}{\bX}
\safemath{\matY}{\bY}
\safemath{\matZ}{\bZ}
\safemath{\matzero}{\bmzero}

\safemath{\matDelta}{\bDelta}
\safemath{\matLambda}{\bLambda}
\safemath{\matPhi}{\bPhi}
\safemath{\matSigma}{\bSigma}
\safemath{\matOmega}{\bOmega}
\safemath{\matTheta}{\bTheta}

\safemath{\matidentity}{\matI}
\safemath{\matone}{\matO}

\safemath{\rnda}{A}
\safemath{\rndb}{B}
\safemath{\rndc}{C}
\safemath{\rndd}{D}
\safemath{\rnde}{E}
\safemath{\rndf}{F}
\safemath{\rndg}{G}
\safemath{\rndh}{H}
\safemath{\rndi}{I}
\safemath{\rndj}{J}
\safemath{\rndk}{K}
\safemath{\rndl}{L}
\safemath{\rndm}{M}
\safemath{\rndn}{N}
\safemath{\rndo}{O}
\safemath{\rndp}{P}
\safemath{\rndq}{Q}
\safemath{\rndr}{R}
\safemath{\rnds}{S}
\safemath{\rndt}{T}
\safemath{\rndu}{U}
\safemath{\rndv}{V}
\safemath{\rndw}{W}
\safemath{\rndx}{X}
\safemath{\rndy}{Y}
\safemath{\rndz}{Z}

\safemath{\rveca}{\bimA}
\safemath{\rvecb}{\bimB}
\safemath{\rvecc}{\bimC}
\safemath{\rvecd}{\bimD}
\safemath{\rvece}{\bimE}
\safemath{\rvecf}{\bimF}
\safemath{\rvecg}{\bimG}
\safemath{\rvech}{\bimH}
\safemath{\rveci}{\bimI}
\safemath{\rvecj}{\bimJ}
\safemath{\rveck}{\bimK}
\safemath{\rvecl}{\bimL}
\safemath{\rvecm}{\bimM}
\safemath{\rvecn}{\bimN}
\safemath{\rveco}{\bomO}
\safemath{\rvecp}{\bimP}
\safemath{\rvecq}{\bimQ}
\safemath{\rvecr}{\bimR}
\safemath{\rvecs}{\bimS}
\safemath{\rvect}{\bimT}
\safemath{\rvecu}{\bimU}
\safemath{\rvecv}{\bimV}
\safemath{\rvecw}{\bimW}
\safemath{\rvecx}{\bimX}
\safemath{\rvecy}{\bimY}
\safemath{\rvecz}{\bimZ}

\safemath{\rvecxi}{\bmxi}
\safemath{\rveclambda}{\bmlambda}
\safemath{\rvecmu}{\bmmu}
\safemath{\rvectheta}{\bmtheta}
\safemath{\rvecphi}{\bmphi}

\safemath{\rmatA}{\bimA}
\safemath{\rmatB}{\bimB}
\safemath{\rmatC}{\bimC}
\safemath{\rmatD}{\bimD}
\safemath{\rmatE}{\bimE}
\safemath{\rmatF}{\bimF}
\safemath{\rmatG}{\bimG}
\safemath{\rmatH}{\bimH}
\safemath{\rmatI}{\bimI}
\safemath{\rmatJ}{\bimJ}
\safemath{\rmatK}{\bimK}
\safemath{\rmatL}{\bimL}
\safemath{\rmatM}{\bimM}
\safemath{\rmatN}{\bimN}
\safemath{\rmatO}{\bimO}
\safemath{\rmatP}{\bimP}
\safemath{\rmatQ}{\bimQ}
\safemath{\rmatR}{\bimR}
\safemath{\rmatS}{\bimS}
\safemath{\rmatT}{\bimT}
\safemath{\rmatU}{\bimU}
\safemath{\rmatV}{\bimV}
\safemath{\rmatW}{\bimW}
\safemath{\rmatX}{\bimX}
\safemath{\rmatY}{\bimY}
\safemath{\rmatZ}{\bimZ}

\safemath{\rmatDelta}{\bimDelta}
\safemath{\rmatLambda}{\bimLambda}
\safemath{\rmatPhi}{\bimPhi}
\safemath{\rmatSigma}{\bimSigma}
\safemath{\rmatOmega}{\bimOmega}
\safemath{\rmatTheta}{\bimTheta}

\usepackage{amssymb}
\usepackage{amsfonts}
\usepackage{mathrsfs}
\usepackage{xspace}
\usepackage{bm}
\usepackage{fancyref}
\usepackage{textcomp}

\usepackage{multirow}
\usepackage{stmaryrd}

\newenvironment{textbmatrix}{	\setlength{\arraycolsep}{2.5pt}%
								\big[\begin{matrix}}{\end{matrix}\big]%
								\raisebox{0.08ex}{\vphantom{M}}}

\def\be{\begin{equation}}
\def\ee{\end{equation}}
\def\een{\nonumber \end{equation}}
\def\mat{\begin{bmatrix}}
\def\emat{\end{bmatrix}}
\def\btm{\begin{textbmatrix}}
\def\etm{\end{textbmatrix}}

\def\ba#1\ea{\begin{align}#1\end{align}}
\def\bas#1\eas{\begin{align*}#1\end{align*}}
\def\bs#1\es{\begin{split}#1\end{split}}
\def\bg#1\eg{\begin{gather}#1\end{gather}}
\def\bml#1\eml{\begin{multline}#1\end{multline}}
\def\bi#1\ei{\begin{itemize}#1\end{itemize}}

\newcommand{\lefto}{\mathopen{}\left}

\DeclareMathOperator{\sign}{sign}			
\DeclareMathOperator*{\argmax}{arg\;max}		
\DeclareMathOperator{\Exop}{\opE}			

\newcommand{\Ex}[2]{\ensuremath{\Exop_{#1}\lefto[#2\right]}} 	
\newcommand{\abs}[1]{\lefto\lvert#1\right\rvert}		

\newcommand{\vecnorm}[1]{\lefto\lVert#1\right\rVert}		

\safemath{\dirac}{\delta}					
\safemath{\krond}{\dirac}					

\safemath{\upto}{\uparrow}
\safemath{\downto}{\downarrow}
\safemath{\iu}{j}							
\safemath{\ev}{\lambda}						
\safemath{\hilseqspace}{l^{2}}				
\newcommand{\banachfunspace}[1]{\setL^{#1}}	
\safemath{\hilfunspace}{\banachfunspace{2}}	
\newcommand{\floor}[1]{\lfloor #1 \rfloor}

\safemath{\SNR}{\textit{SNR}} 				
\safemath{\PAR}{\textit{PAR}} 				
\safemath{\No}{N_0}							
\safemath{\Es}{E_s}							
\safemath{\Eb}{E_b}							
\safemath{\EbNo}{\frac{\Eb}{\No}}
\safemath{\EsNo}{\frac{\Es}{\No}}

\DeclareMathOperator{\CHop}{\ensuremath{\opH}} 
\safemath{\tvir}{\rndh_{\CHop}}				
\safemath{\tvtf}{\rndl_{\CHop}}				
\safemath{\spf}{\rnds_{\CHop}}				
\safemath{\bff}{H_{\CHop}}					

\safemath{\ircf}{r_{h}}						
\safemath{\tftvcf}{r_{s}}					
\safemath{\tfcf}{r_{l}}						
\safemath{\bfcf}{r_{H}}						

\safemath{\tcorr}{c_h}						
\safemath{\scf}{c_{s}}						
\safemath{\tfcorr}{c_{l}}					
\safemath{\fcorr}{c_{H}}						

\safemath{\mi}{I}							
\safemath{\capacity}{C}						

\safemath{\normal}{\mathcal{N}}			
\safemath{\jpg}{\mathcal{CN}}			
\safemath{\mchain}{\leftrightarrow}		

\safemath{\dB}{\,\mathrm{dB}}
\safemath{\dBm}{\,\mathrm{dBm}}
\safemath{\Hz}{\,\mathrm{Hz}}
\safemath{\kHz}{\,\mathrm{kHz}}
\safemath{\MHz}{\,\mathrm{MHz}}
\safemath{\GHz}{\,\mathrm{GHz}}
\safemath{\s}{\,\mathrm{s}}
\safemath{\ms}{\,\mathrm{ms}}
\safemath{\mus}{\,\mathrm{\text{\textmu}s}}
\safemath{\ns}{\,\mathrm{ns}}
\safemath{\ps}{\,\mathrm{ps}}
\safemath{\meter}{\,\mathrm{m}}
\safemath{\mm}{\,\mathrm{mm}}
\safemath{\cm}{\,\mathrm{cm}}
\safemath{\m}{\,\mathrm{m}}
\safemath{\W}{\,\mathrm{W}}
\safemath{\mW}{\, \mathrm{mW}}
\safemath{\J}{\,\mathrm{J}}
\safemath{\K}{\,\mathrm{K}}
\safemath{\bit}{\,\mathrm{bit}}
\safemath{\nat}{\,\mathrm{nat}}

\safemath{\define}{\triangleq}			

\safemath{\equivalent}{\sim}
\safemath{\distas}{\sim}					
\safemath{\sdiff}{\Delta}				

\safemath{\reals}{\mathbb{R}}
\safemath{\positivereals}{\reals_{+}}
\safemath{\integers}{\mathbb{Z}}
\safemath{\posint}{\integers_{+}}
\safemath{\naturals}{\mathbb{N}}
\safemath{\posnaturals}{\naturals_{+}}
\safemath{\complexset}{\mathbb{C}}
\safemath{\rationals}{\mathbb{Q}}

\newcommand*{\fancyrefapplabelprefix}{app}		
\newcommand*{\fancyrefthmlabelprefix}{thm}		
\newcommand*{\fancyreflemlabelprefix}{lem}		
\newcommand*{\fancyrefcorlabelprefix}{cor}		
\newcommand*{\fancyrefdeflabelprefix}{def}		
\newcommand*{\fancyrefproplabelprefix}{prop}		
\newcommand*{\fancyrefexmpllabelprefix}{exmpl}
\newcommand*{\fancyrefalglabelprefix}{alg}		
\newcommand*{\fancyreftbllabelprefix}{tbl}		

\frefformat{vario}{\fancyrefseclabelprefix}{Section~#1}
\frefformat{vario}{\fancyrefthmlabelprefix}{Theorem~#1}
\frefformat{vario}{\fancyreftbllabelprefix}{Table~#1}
\frefformat{vario}{\fancyreflemlabelprefix}{Lemma~#1}
\frefformat{vario}{\fancyrefcorlabelprefix}{Corollary~#1}
\frefformat{vario}{\fancyrefdeflabelprefix}{Definition~#1}
\frefformat{vario}{\fancyreffiglabelprefix}{Fig.~#1}
\frefformat{vario}{\fancyrefapplabelprefix}{Appendix~#1}
\frefformat{vario}{\fancyrefeqlabelprefix}{(#1)}
\frefformat{vario}{\fancyrefproplabelprefix}{Proposition~#1}
\frefformat{vario}{\fancyrefexmpllabelprefix}{Example~#1}
\frefformat{vario}{\fancyrefalglabelprefix}{Algorithm~#1}

 \newtheorem{thm}{Theorem}
 
 \newtheorem{defi}{Definition}
 \newtheorem{lem}[thm]{Lemma}
 
\safemath{\dictab}{[\,\dicta\,\,\dictb\,]}

\safemath{\ysig}{\bmy}
\safemath{\ysighat}{\hat{\ysig}}
\safemath{\ysigdim}{M}
\safemath{\xsig}{\bmx}
\safemath{\xsigdim}{N}
\safemath{\nx}{n_x}
\safemath{\zsig}{\bmz}
\safemath{\zsigdim}{\ysigdim}
\safemath{\rsig}{\bmr}
\safemath{\Adict}{\bA}
\safemath{\Adicttilde}{\widetilde{\Adict}}
\safemath{\Adictdim}{\outputdim\times\xsigdim}
\safemath{\avec}{\bma}
\safemath{\avectilde}{\tilde{\avec}}
\safemath{\Bdict}{\bB}
\safemath{\Bdicttilde}{\widetilde{\Bdict}}
\safemath{\Cdict}{\bC}
\safemath{\cvec}{\bmc}
\safemath{\Ddict}{\bD}
\safemath{\Ddictdim}{\ysigdim\times\xsigdim}
\safemath{\dvec}{\bmd}
\safemath{\Ddicttilde}{\widetilde{\bD}}
\safemath{\Bonb}{\bB}
\safemath{\bvec}{\bmb}
\safemath{\Bonbdim}{\ysigdim\times\ysigdim}
\safemath{\noise}{\bmn}
\safemath{\noisedim}{\ysigim}
\safemath{\err}{\bme}
\safemath{\errdim}{\ysigdim}
\safemath{\errset}{\setE}
\safemath{\nerr}{n_e}
\safemath{\delop}{\bP_\errset}
\safemath{\delopc}{\bP_{{\errset}^c}}

\safemath{\cplxi}{\imath}
\safemath{\cplxj}{\jmath}

\safemath{\dict}{\matD}
\safemath{\inputdim}{N}		
\safemath{\outputdim}{M}		
\safemath{\sparsity}{S}	
\safemath{\inputdimA}{{N_a}}	
\safemath{\inputdimB}{{N_b}}	
\safemath{\elemA}{{n_a}}	
\safemath{\elemB}{{n_b}}	
\safemath{\resA}{\matR_a}	
\safemath{\resB}{\matR_b}	
\safemath{\subD}{\matS} 
\safemath{\subA}{\matS_a} 
\safemath{\subB}{\matS_b} 
\safemath{\dicta}{\matA} 	
\safemath{\dictb}{\matB} 	
\safemath{\hollowS}{H}
\safemath{\hollowA}{H_a}
\safemath{\hollowB}{H_b}
\safemath{\cross}{Z}
\safemath{\coh}{\mu_d}			
\safemath{\coha}{\mu_a}			
\safemath{\cohb}{\mu_b}			
\safemath{\mubs}{\nu}	
\safemath{\cohm}{\mu_m} 
\safemath{\dictset}{\setD}	
\safemath{\dictsetp}{\dictset(\coh,\coha,\cohb)}	
\safemath{\dictsetgen}{\dictset_\text{gen}}
\safemath{\dictsetgenp}{\dictsetgen(\coh)}
\safemath{\dictsetonb}{\dictset_\text{onb}}
\safemath{\dictsetonbp}{\dictsetonb(\coh)}

\safemath{\leftside}{U}
\safemath{\rightsideA}{R_a}
\safemath{\rightsideB}{R_b}

\safemath{\indexS}{\setI_S} 

\safemath{\na}{n_a}			
\safemath{\nb}{n_b}			
\safemath{\coeffa}{p_i}	
\safemath{\coeffb}{q_j}	
\safemath{\seta}{\setP}		
\safemath{\setb}{\setQ}     
\safemath{\setw}{\setW}	
\safemath{\setz}{\setZ}	
\safemath{\cola}{\veca}		
\safemath{\colb}{\vecb}		
\safemath{\cold}{\vecd}		
\safemath{\inputvec}{\vecx} 	
\safemath{\error}{\vece}	
\safemath{\noiseout}{\vecz} 	
\safemath{\inputvecel}{x}
\safemath{\inputveca}{\vecx_a}
\safemath{\inputvecb}{\vecx_b}
\safemath{\outputvec}{\vecy}	
\safemath{\lambdamin}{\lambda_{\mathrm{min}}}

\safemath{\elltwo}{\ell_2}
\safemath{\ellone}{\ell_1}
\safemath{\ellzero}{\ell_0}
\safemath{\ellinf}{\ell_\infty}
\safemath{\ellinftilde}{\ell_{\widetilde\infty}}
\safemath{\licard}{Z(\coh,\coha,\cohb)}
\safemath{\xsol}{\hat{x}}
\safemath{\xbord}{x_b}		
\safemath{\xstat}{x_s}		
\safemath{\xstatLone}{\tilde{x}_s}
\safemath{\order}{\mathcal{O}} 
\safemath{\scales}{\Theta} 
\safemath{\ones}{\mathbf{1}} 
\safemath{\zeroes}{\mathbf{0}} 
\safemath{\thlone}{\kappa(\coh,\cohb)} 
\safemath{\constoneA}{\delta} 
\safemath{\constoneB}{\epsilon} 
\safemath{\nlarge}{L}				   
\safemath{\sumlarge}{S_\nlarge}
\safemath{\maxlarger}{P_\nlarge}	   
\safemath{\Pzero}{\textrm{P0}}	
\safemath{\Pone}{\textrm{P1}}
\safemath{\vecfir}{\vecw}			 
\safemath{\vecsec}{\vecz}
\safemath{\elvecfir}{w}              
\safemath{\elvecsec}{z}				 
\safemath{\nlargefir}{n}
\safemath{\normout}{\gamma}
\safemath{\auxfun}{h}
\safemath{\supp}{\textrm{supp}}

\safemath{\indexa}{\ell}
\safemath{\indexb}{r}
\safemath{\indexc}{i}
\safemath{\indexd}{j}

\safemath{\project}{P}

\usepackage{framed}

\IEEEoverridecommandlockouts
\allowdisplaybreaks 

\newcommand{\norm}[1]{\left\lVert #1 \right\rVert}

\newcommand{\PR}[1]{\ensuremath{\!\left[#1\right]}}
\newcommand{\PC}[1]{\ensuremath{\!\left(#1\right)}}
\newcommand{\chav}[1]{\ensuremath{\!\left\{#1\right\}}}

\safemath{\Hj}{\bmj}
\safemath{\sj}{w}
\safemath{\Ej}{E_w}
\safemath{\quant}{Q}
\safemath{\compquant}{\mathcal{Q}}
\safemath{\Cy}{\bC_{\bmy}}

\begin{document}

\title{High Dynamic Range mmWave Massive MU-MIMO with Householder Reflections}
\author{\IEEEauthorblockN{Victoria Palhares, Gian Marti, Oscar Casta\~neda, and Christoph Studer}\\
\em Department of Information Technology and Electrical Engineering, ETH Zurich, Switzerland \\ e-mail: vmenescal@ethz.ch, gimarti@ethz.ch, caoscar@ethz.ch, and studer@ethz.ch\\
\thanks{This work was supported in part by an ETH Research Grant and by the Swiss National Science Foundation (NSF). The work of CS was supported in part by the U.S. NSF under grants CNS-1717559 and ECCS-1824379.}
\thanks{We thank Remcom for providing a license for Wireless InSite~\cite{Remcom}.}
}

\maketitle


\begin{abstract}
All-digital massive multiuser (MU) multiple-input multiple-output (MIMO) at millimeter-wave (mmWave) frequencies is a promising technology for next-generation wireless systems. Low-resolution analog-to-digital converters (ADCs) can be utilized to reduce the power consumption of all-digital basestation (BS) designs. However, simultaneously transmitting user equipments (UEs) with vastly different BS-side receive powers either drown weak UEs in quantization noise or saturate the ADCs. To address this issue, we propose \emph{high dynamic range (HDR) MIMO}, a new paradigm that enables simultaneous reception of strong and weak UEs with low-resolution ADCs. HDR MIMO combines an adaptive analog spatial transform with digital equalization: The spatial transform focuses strong UEs on a subset of ADCs in order to mitigate quantization and saturation artifacts; digital equalization is then used for data detection. We demonstrate the efficacy of HDR MIMO in a massive MU-MIMO mmWave scenario that uses Householder reflections as spatial transform.
\end{abstract}


\section{Introduction}

Next-generation wireless systems are expected to combine communication at millimeter-wave (mmWave) frequencies with massive multiuser (MU) multiple-input multiple-output (MIMO) technology~\cite{Swindlehurst2014}. While communication at mmWave frequencies provides access to large, contiguous, and available portions of the frequency spectrum, 
massive MU-MIMO provides beamforming gains that combat the high path-loss at mmWave frequencies and enables  communication with multiple user equipments (UEs) in the same time-frequency resource.

The deployment of all-digital massive MIMO basestations (BSs), i.e.,~BSs where each antenna has a dedicated radio-frequency (RF) chain and dedicated data converters, comes at the cost of increased digital circuit complexity, power consumption, and interconnect data rates~\cite{Dutta2020,Mirfarshbafan2020}. To alleviate these drawbacks, one can use low-resolution analog-to-digital converters (ADCs) in the uplink~\cite{Jacobsson2017}. However, low-resolution ADCs do not quantize high dynamic range signals properly. In a scenario where one UE is significantly stronger than the other UEs, weak UE signals either drown in quantization noise, if the BS adjusts the ADC input gains to the strongest UE, or the receive signals are distorted due to ADC saturation caused by the strongest UE, if the BS adjusts the ADC input gains to the weakest UEs. Since quantization and saturation effects are, in general, irreversible, digital equalization cannot fully mitigate such distortions \cite{Geiger2011}. 

\subsection{Contributions}

We propose \textit{high dynamic range (HDR) MIMO}, a new paradigm that deals with receive signals of high dynamic range in all-digital BS architectures that are equipped with low-resolution ADCs. HDR MIMO utilizes one or multiple adaptive analog spatial transforms that focus either the energy of the strongest UE or the signal dimension with the highest receive power on a pair of ADCs (for the in-phase and quadrature components) per transform; digital equalization is then used to detect the transmitted data of the strong UE as well as of the weak UEs. For the adaptive analog spatial transforms, we propose two distinct Householder reflections (HRs): (i)~strongest UE isolation (HR-ISO), which isolates the energy of the strongest UE on dedicated ADCs, and (ii)~maximum power isolation (HR-MAX), which focuses the signal dimension with the highest receive power on dedicated ADCs. We demonstrate the efficacy of HDR MIMO by simulating a mmWave massive MU-MIMO uplink system with an  all-digital BS that utilizes low-resolution ADCs and channel vectors from a commercial ray tracer.

\subsection{Related Work}

Past work on interference or jammer mitigation has led to the development of methods that perform spatial nulling of strong signals; see, e.g., \cite{Subbaram1993,Yan2014,Marti2021a,Marti2021b}. For example, spatial projection onto the orthogonal subspace (POS, for short) of the jammer channel has been proposed for jammer mitigation in~\cite{Yan2014,Subbaram1993}. Since the content of the interference or jamming signal is typically of no interest, it can be nulled and discarded. In contrast, HDR MIMO assumes that all signal sources are of interest and enables the simultaneous detection of weak and strong received signals. Furthermore, if one is interested in analyzing interferer or jammer signals in the context of jammer mitigation, then HDR MIMO provides means to recover such signals jointly with detecting the legitimate UE~data. 

The use of analog transforms to perform spatial filtering for systems with low-resolution BS designs has been proposed in~\cite{Marti2021a,Marti2021b} for jammer mitigation, and in \cite{Alaei2021} for strong adjacent channel interference mitigation. In \cite{Marti2021a}, the authors propose HERMIT, which uses an adaptive analog transform to remove most of the jammer's energy prior to sampling by the ADCs. In~\cite{Marti2021b}, the authors propose beam-slicing, a nonadaptive analog transform that focuses the jammer energy onto few ADCs. In~\cite{Alaei2021}, the authors propose a hybrid beamformer with spatial filtering that adaptively mitigates interference. In contrast to these methods, we propose the use of adaptive analog spatial HRs that focus the energy of the strongest UE or the signal dimension with the highest receive power onto a subset of the available ADCs without discarding any signals, which enables the simultaneous recovery of signals from strong and weak UEs. 

Modulo ADCs that rely on a concept known as unlimited sampling~\cite{Bhandari2021,Bhandari2022} have been proposed in~\cite{Liu2023} to develop massive MIMO receiver architectures that are able to deal with high dynamic range signals. However, unlimited sampling requires specialized signal recovery techniques to reconstruct the high dynamic range signals. Furthermore, existing modulo ADC designs do not yet achieve the sampling rates required by modern wireless systems. In contrast, our proposed methods build upon conventional massive MIMO architectures with standard automatic gain controllers (AGCs) and  ADCs, which can support the bandwidth of modern wireless systems and do not need specialized signal recovery methods. 

In MU communication, one of the challenges is the near-far problem, i.e., if a UE is physically much closer to the BS than another UE, then the BS-side receive power of the close UE is often much stronger than {that of the other UEs}. Power control at the UE side is typically used to confine the dynamic range at the receiver side~\cite{Tse2005,Chiang2008,Saxena2015,Bjornson2017}, reducing the necessity of HDR MIMO. If, however, there exist UEs, rogue transmitters, interferers, or jammers in the same or nearby frequency bands that do \emph{not} adhere to such power control mechanisms, then HDR MIMO still enables reliable data transmission.   

\subsection{Notation}
Upper case and lower case bold symbols denote matrices and column vectors, respectively; uppercase calligraphic letters denote sets. We use ~$\PR{\bma}_i$ for the $i$th element of vector $\bma$, $\PR{\bA}_{\PC{i,j}}$ for the element in the $i$th row and $j$th column of $\bA$, and~$\bma_j$ for the $j$th column of~$\bA$. The $M \times M$ identity matrix is~$\bI_M$ and the $M$-dimensional unit vector is~$\bme_i$, where the $i$th entry is one and the rest is zero. The superscripts~$(\cdot)^\mathrm{*}$, $(\cdot)^\mathrm{T}$, and~$(\cdot)^\mathrm{H}$ denote the complex conjugate, transpose, and Hermitian transpose, respectively. A diagonal matrix with the vector~$\bma$ on its main diagonal is $\mathrm{diag} \PC{\bma}$. The Euclidean norm of $\bma$ is $\vecnorm{\bma}$. The eigenvalue decomposition of a symmetric matrix $\bA \in \opC^{M \times M}$ is $\bA = \bL \boldsymbol\Lambda \bL^H = \sum_{m=1}^M \lambda_m\boldsymbol{\ell}_m \boldsymbol{\ell}_m^H$, where~$\lambda_m$, $m=1,\ldots,M$, are the eigenvalues, which are sorted in descending order, and~$\boldsymbol{\ell}_m$ are the corresponding unit-length eigenvectors. The operators $\Re\chav{a}$ and $\Im\chav{a}$ extract the real and imaginary part of $a\in\complexset$, respectively. The complex-valued sign (or phase) of  $a\in\complexset$ is $\sign(a) = {a}/\abs{a}$, which we define to be $1$ if $a=0$. The floor function $\floor{a}$ yields the greatest integer less than or equal to $a\in\opR$. The expectation operator is $\Ex{}{\cdot}$. We use $j$ as the imaginary unit.


\section{Prerequisites}

\subsection{System Model}
\label{sec:system_model}

We consider the uplink of a mmWave massive MU-MIMO system with $U$ single-antenna UEs transmitting data to an all-digital BS with $B$ antennas. We consider frequency-flat channels with the following baseband input-output relation:
\begin{align}
\label{eq:received_vector}
\bmy  = \bG \bD \bms + \bmn. 
\end{align}
Here,  $\bmy \in\opC^{B}$ is the unquantized BS-side receive vector, $\bG \in \opC^{B \times U}$ is the uplink channel matrix, $\bD = \mathrm{diag} \PC{d_1,\dots,d_{U}}\in \opR^{U \times U}$ is the power control matrix, $\bms \in \setX^{U}$ is the transmit symbol vector with entries taken from a constellation $\setX$, and $\bmn \in \opC^{B}$ models noise as an i.i.d.\ circularly-symmetric complex Gaussian random vector with variance~$\No$ per entry. The constellation $\setX$ is normalized so that $\Ex{}{|[\bms]_u|^2} = 1$, $u=1,\ldots,U$. We define the effective channel matrix as $\bH = \bG \bD$. We assume that the columns of $\bH$ are sorted in descending order with respect to the column norms $\norm{\bmh_u}$, $u=1,\dots,U$. The matrices $\bG$ and $\bD$ are sorted accordingly. 

In what follows, we consider a scenario in which the {BS-side} receive power of the first UE is much stronger than that of the other UEs. We model such a situation by assuming that all UEs except for the first UE adhere to power control, which is expressed through the per-UE gains $d_2,\dots,d_U$ of the power control matrix $\bD$ (cf.~\fref{sec:simulated_scenario} for the details on the used power control strategy). For the first UE, the gain~$d_1$ can be expressed by the dynamic range $\rho$ (in decibels) of the BS-side receive power between the strongest and weakest UEs as
\begin{align}\label{eq:rho_definition}
\rho = 10 \log_{10} \PC{\frac{d_1^2 \| \bmg_1 \|^2}{d_U^2 \| \bmg_U\|^2}}\!. 
\end{align}
As a result, the squared Euclidean norm of the strongest UE's channel vector $\bmh_1$ is $\rho$ decibels larger than the squared norm of the weakest UE's channel vector $\bmh_U$. Here, $\bmh_1$ and $\bmh_U$ correspond to the first and last columns of the effective channel matrix $\bH$, respectively. 

To estimate the channel matrix $\bH$, we assume block fading where the UEs transmit orthogonal pilots $\bS_T \in \opC^{U \times K}$ for the duration of $K \geq U$ time slots. Each column of~$\bS_T$ contains the pilots of all UEs transmitted during one training time slot. The (unquantized) receive vectors in the training phase are modeled as 
$\bY_T = \bH \bS_T + \bN_T$, where each column in the receive matrix $\bY_T \in \opC^{B \times K}$ and the noise matrix $\bN_T \in \opC^{B \times K}$ corresponds to one training time slot. We estimate the effective channel matrix~$\bH$ using the least-squares estimator $\hat{\bH} = \bY_T \bS_T^{\mathrm{H}} \PC{\bS_T \bS_T^{\mathrm{H}}}^{-1}$. In what follows, we define $\underline{\bmh} = \bmh_1$ as the channel vector associated with the first (strongest) UE and $\underline{\hat{\bmh}}$ as its estimate.

\begin{figure}[tp]
\centering
\includegraphics[width=0.99\columnwidth]{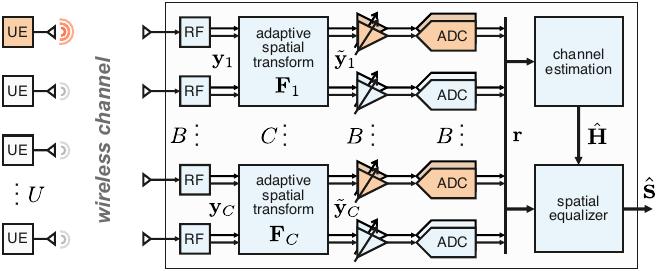}
\caption{HDR MIMO receiver architecture: The RF chains, adaptive analog spatial transforms $\bF_c$, $c=1,\ldots,C$, AGCs, and 
ADCs are divided into $C$ antenna clusters. In a scenario in which the BS-side receive power of one UE (highlighted in orange) is much stronger than that of the other UEs, each adaptive analog spatial transform focuses the energy of the strongest UE on a pair of AGCs and ADCs per cluster (highlighted in orange); this prevents saturation of the remaining ADCs in the same cluster. The transmitted UE data can then be recovered by a spatial equalizer.} \label{fig:schematic}
\end{figure}

\subsection{High Dynamic Range (HDR) MIMO Receiver} \label{sec:hdr_mimo_receiver}

We now describe the HDR MIMO BS architecture depicted in \fref{fig:schematic}, which is able to deal with receive signals of high dynamic range. In order to prevent weak UEs from drowning in quantization noise or strong UEs from saturating all ADCs, we introduce adaptive analog spatial transforms prior to the AGCs and ADCs. We focus on linear transforms represented by the matrices $\bF_c \in \opC^{S \times S}$,  $c=1,\ldots,C$, where each matrix operates on a cluster of $S=B/C$ antennas with $C$ such antenna clusters in total.\footnote{In practice, the number $S$ of antennas per cluster should not be too large as implementing large analog spatial transforms is challenging \cite{Enciso2020}.} These {adaptive} analog spatial transforms implement 
\begin{align} \label{eq:y_tilde_c}
\tilde{\bmy}_c = \bF_c \bmy_c,~c=1,\dots,C,
\end{align}
where $\bmy_c \in \opC^{S}$ is the $c$th cluster of the receive vector $\bmy = \PR{\bmy_1^\mathrm{T},\dots,\bmy^\mathrm{T}_C}^\mathrm{T} \in \opC^B$ and $\tilde\bmy_c \in \opC^{S}$ is the corresponding output of the adaptive analog spatial transform. The transformed receive vector is, therefore,
\begin{align} \label{eq:y_tilde}
\tilde{\bmy} = \bF \bmy,
\end{align}
where $\tilde{\bmy} = \PR{\tilde{\bmy}_1^\mathrm{T},\dots,\tilde{\bmy}^\mathrm{T}_C}^\mathrm{T} \in \opC^B$ and $\bF = \mathrm{diag} \PC{\bF_1,\dots,\bF_C}$ is a $B \times B$ block diagonal matrix.

Note that if no structural constraints are imposed on the clusters of the {adaptive analog} spatial transform matrix $\bF$, then~\fref{eq:y_tilde} would have to be computed as a matrix-vector product
consisting of $B^2/C$ {complex-valued} multiplications. However, the specific structure of the per-cluster transforms~$\bF_c$,  $c=1,\ldots,C$, proposed in~\fref{sec:analogtransforms}, enables simpler analog hardware with fewer complex-valued multiplications. 

\subsection{Analog-to-Digital Conversion} \label{sec:quantization}

As we will show in \fref{sec:analogtransforms}, the adaptive analog spatial transform ensures that only a subset of its outputs contains strong signals. One can then utilize the AGCs  (cf.~\fref{fig:schematic}) to individually adjust the input gains for each pair of ADCs with the goal of minimizing saturation and quantization artifacts. In order to model such ADC artifacts, we follow~\cite{Marti2021a, Marti2021b} and apply a uniform midrise quantizer to each entry of the real and imaginary parts of the transformed receive vector~$\tilde{\bmy}$ in \fref{eq:y_tilde}. The scalar function $Q(\cdot)$ that models each ADC is given by
\begin{equation}
Q(x) = 
\begin{cases} \label{eq:uniform_midrise_quantizer}
\Delta \floor{\frac{x}{\Delta}}+\frac{\Delta}{2}, & \mathrm{if}\ |x| \leq \Delta 2^{q-1},\\
\frac{\Delta}{2}(2^q-1)\frac{x}{|x|}, & \mathrm{if}\ |x| >\Delta 2^{q-1},
\end{cases}
\end{equation}
where $\Delta$ is the quantizer's step size and $q$ is the number of quantization bits. The distortions introduced by this model are then taken into account by the spatial equalizer (cf.~\fref{fig:schematic}).

To design the spatial equalizer, we employ Bussgang's decomposition~\cite{Bussgang1952}, where the effect of the quantizer on a real-valued zero-mean random variable (RV) $x$ is modeled as
\begin{align} \label{eq:bussgang_decomposition}
	Q(x) = \gamma x + d,
\end{align}
with the so-called Bussgang gain 
\begin{align} \label{eq:buggang_gain}
	\gamma = \frac{\Ex{}{Q(x) x}}{\Ex{}{x^2}}. 
\end{align}
In \fref{eq:bussgang_decomposition}, the choice of $\gamma$ as in \fref{eq:buggang_gain} ensures that the distortion $d$ is uncorrelated with $x$ and is zero mean with variance
\begin{align} \label{eq:D_definition}
D = \Ex{}{d^2} = \Ex{}{\quant(x)^2} - \gamma^2 \Ex{}{x^2}\!.
\end{align}
In order to minimize the distortion caused by saturation and quantization, one can either pick an optimal step size $\Delta$ while fixing the input variance or pick an optimal input variance while fixing the step size. We force the variance of the input~$x$ to one and select an optimal step size~$\Delta$ that minimizes the mean squared error (MSE) between the quantizer's output~$Q(x)$ and its input $x$ \cite{Max1960}.
To enforce unit variance at the inputs of all ADCs, we need to adjust the AGC settings, which we model by a diagonal gain control matrix $\boldsymbol{\Omega} =\mathrm{diag}\PC{\omega_1,\dots,\omega_B}$ with  
\begin{align} \label{eq:ADC_gains}
\omega_b = \sqrt{\frac{2}{\PR{\bC_{\tilde{\bmy}}}_{\ \PC{b,b}}}},  \quad b=1,\ldots,B.
\end{align}
Here, the covariance matrix $\bC_{\tilde{\bmy}}$ is given by
\begin{align}\label{eq:true_cy}
\bC_{\tilde{\bmy}} &= \Ex{}{\tilde{\bmy} \tilde{\bmy}^{\mathrm{H}}} = \bF \Ex{}{\bmy \bmy^{\mathrm{H}}} \bF^{\mathrm{H}}  = \bF \bC_{\bmy}\bF^{\mathrm{H}}.
\end{align}
By multiplying the gain control matrix $\boldsymbol{\Omega}$ to the transformed receive vector  $\tilde{\bmy}$,  the entries of $\boldsymbol{\Omega} \tilde{\bmy}$ have unit variance per real dimension. The resulting AGC and ADC model is  
\begin{align} \label{eq:quantized_y_tilde}
\bmr &= \compquant(\tilde{\bmy}) = Q \PC{\Re \chav{\boldsymbol{\Omega} \tilde{\bmy}}} +  j Q \PC{\Im \chav{\boldsymbol{\Omega} \tilde{\bmy}}}.
\end{align}
By applying Bussgang's decomposition \fref{eq:bussgang_decomposition} individually to each entry of~\fref{eq:quantized_y_tilde}, we obtain the following linearized input-output relation:
\begin{align}
	\bmr &= Q \PC{\Re \chav{\boldsymbol{\Omega} \tilde{\bmy}}} +  j Q \PC{\Im \chav{\boldsymbol{\Omega} \tilde{\bmy}}}\\
	&= \gamma \Re \chav{\boldsymbol{\Omega} \tilde{\bmy}} + \bmd_r +  j \PC{\gamma \Im \chav{\boldsymbol{\Omega} \tilde{\bmy}} + \bmd_j}\\
	&= \gamma \boldsymbol{\Omega}\PC{\Re \chav{ \tilde{\bmy}} + j \Im\chav{ \tilde{\bmy}}} + \bmd\
	= \gamma \boldsymbol{\Omega} \tilde{\bmy}  + \bmd.  
	\label{eq:quantized_y_tilde_2}
\end{align}	
Here, we define $\bmd = \bmd_r + j \bmd_j$, where $\bmd_r$ and $\bmd_j$ are the real and imaginary parts of the distortions caused by $\compquant(\tilde{\bmy})$,~respectively.

\subsection{Spatial Equalization} 

The linearized input-output relation in \fref{eq:quantized_y_tilde_2} can now be used to design a linear {minimum mean squared error} (LMMSE)-type equalizer, which generates estimates for the transmitted symbols as $\hat{\bms} = \bW \bmr$. In the equalization step, we do not assume perfect knowledge of the channel matrix. Therefore, using the estimate $\hat{\bH}$ of the channel matrix $\bH$, we define our LMMSE-type equalization matrix as
\begin{align} \label{eq:lmmse_equalization_matrix}
\bW = \frac{1}{\gamma} \hat\bH^{\mathrm{H}} \bF^{\mathrm{H}} \boldsymbol{\Omega} \bigg(&  \boldsymbol{\Omega} \bF \hat\bH \hat\bH^{\mathrm{H}}\bF^{\mathrm{H}} \boldsymbol{\Omega} \notag \\
 &  + \No \boldsymbol{\Omega} \bF \bF^{\mathrm{H}} \boldsymbol{\Omega} + \frac{2D}{\gamma^2} \bI_B  \bigg)^{\!-1}, 
\end{align}
where we approximate the distortion covariance matrix as
\begin{equation}
	\bC_\bmd = \Ex{}{\bmd \bmd^\mathrm{H}} \approx 2D\,\bI_B,  \label{eq:assump_diag} 
\end{equation}
with $D$ from \fref{eq:D_definition}.


\section{Adaptive Analog Spatial Transforms}
\label{sec:analogtransforms}

We are now ready to describe the specifics of suitable adaptive analog spatial transforms~$\bF_c$, $c=1,\ldots,C$, which we use to focus strong receive signals onto a subset of the ADCs. 
Specifically, we present two HR matrices designed for different purposes: HR-ISO, which isolates the energy of the strongest UE on dedicated ADCs, and HR-MAX, which isolates the signal dimension with the highest receive power on dedicated~ADCs.  

\subsection{Householder Reflections}

An HR is carried out by multiplying a vector by a matrix~$\bQ_{\bmv}$ that reflects the vector with respect to a subspace described by its normal vector $\bmv$~\cite{Trefethen1997,Golubl996}. 
\begin{defi}
Let $\bmv \in\complexset^M$ be a nonzero vector. Then, the HR matrix $\bQ_{\bmv}$ is defined as follows:
\begin{align} \label{eq:householder_matrix}
\bQ_{\bmv}= \bI_M - 2 \frac{\bmv \bmv^\mathrm{H}}{\|\bmv\|^2}.
\end{align}
\end{defi}
HR matrices are (i) unitary so that the statistics of an i.i.d Gaussian random vector remain unaffected and (ii) Hermitian symmetric, i.e., $\bQ_{\bmv}^\mathrm{H}\bQ_{\bmv}=\bQ_{\bmv}\bQ_{\bmv}=\bI_M$. 

By replacing $\bF_c$ in \fref{eq:y_tilde_c} with $\bQ_{\bmv_c}$ as in \fref{eq:householder_matrix}, we obtain a transformed receive vector in the $c$th antenna cluster as
\begin{align} \label{eq:transformed_receive_vector_cluster}
\tilde{\bmy}_c = \bQ_{\bmv_c} \bmy_c = \bmy_c - 2 \frac{\bmv_c \bmv_c^\mathrm{H}}{\|\bmv_c\|^2} \bmy_c,
\end{align}
where $\bmv_c \in \opC^S$. This operation requires only one inner-product calculation $\PC{\bmv_c^{\mathrm{H}}/\| \bmv_c\|}\bmy_c$ followed by subtracting a scaled version of  $\bmv_c/\|\bmv_c\|$ from $\bmy_c$. Thus, the number of real- and complex-valued multiplications required to transform all~$C$ {antenna} clusters decreases from $B^2/C$ to only $2B+C$, which can be implemented more efficiently, e.g., using analog multiplication circuitry originally developed for beamforming~\cite{Naviasky2021}. We reiterate that the statistics of the noise vector $\bmn$ in \fref{eq:received_vector} remain unaffected by \fref{eq:transformed_receive_vector_cluster} since the matrices $\bQ_{\bmv_c}$, $c=1,\ldots,C$, are unitary. Suitable choices for the vector $\bmv_c$ are discussed next.

\subsection{Strongest UE Isolation (HR-ISO)} \label{sec:ISO}

As the first choice for the HR matrix, we wish to isolate the energy of the strongest UE on the $i$th output of the {adaptive analog} spatial transform. To this end, we need the following result; the proof is given in \fref{app:ISO}. 
\begin{lem} \label{lem:HouseholderQR}
Let $\bma\in\complexset^M$ be a nonzero vector. Then, the vector
\begin{align}   \label{eq:solutionoptimizationproblem1}
\hat\bmv = \, \bma + \|\bma\| \sign([\bma]_i) \bme_i 
\end{align}
is a solution of the following optimization problem:
\begin{align} \label{eq:optimizationproblem1}
\hat\bmv \in \argmax_{\tilde\bmv\in\complexset^M }\, | \bme_i^\mathrm{H} \bQ_{\tilde\bmv} \bma|^{2}.
\end{align}
\end{lem}

We can use \fref{lem:HouseholderQR} to isolate the energy of the strongest UE on the first pair of AGCs/ADCs of each antenna cluster by setting $i=1$ and $\bma=\underline{\hat{\bmh}}_c$, which corresponds to solving
\begin{align} \label{eq:specificoptimizationproblem1}
\hat{\bmv}^\mathrm{HR-ISO}_c \in \argmax_{\tilde\bmv_c \in\complexset^S }\, |\bme_1^\mathrm{H}  \bQ_{\tilde{\bmv}_c} \underline{\hat{\bmh}}_c|^{2},
\end{align}
where $\underline{\hat{\bmh}}_c = \PR{[\underline{\hat{\bmh}}]_{(c-1)S+1},\dots,[\underline{\hat{\bmh}}]_{cS}}^\mathrm{T} \in \opC^{S}$ is the estimated channel vector associated with the $c$th cluster of the strongest UE. According to \fref{lem:HouseholderQR}, a vector that solves  \fref{eq:specificoptimizationproblem1} is  
\begin{align}
\hat{\bmv}^\mathrm{HR-ISO}_c = \underline{\hat{\bmh}}_c + \| \underline{\hat{\bmh}}_c \| \mathrm{sign}([\underline{\hat{\bmh}}_c]_{1})\bme_1.
\end{align}
Therefore, the HR matrix that focuses {the energy of the strongest UE} on the first per-cluster AGCs and ADCs is~$\bQ_{\hat{\bmv}^\mathrm{HR-ISO}_c}$. Note that $\hat{\bmv}^\mathrm{HR-ISO}_c$ is the standard HR vector used for QR decompositions applied to $\underline{\hat{\bmh}}_c$ \cite[Sec.~5.2.1]{Golubl996}.

\subsection{Maximum Power Isolation (HR-MAX)}
\label{sec:MAX}
As the second choice for the HR matrix, we wish to isolate the signal dimension with the highest receive power per antenna cluster on the $i$th output of the adaptive analog spatial transform. To this end, we need the following result; the proof is given in \fref{app:MAX}.

\begin{lem}\label{lem:maxpowerisolation}
Let $\bma\in\complexset^M$ be a zero-mean random vector with covariance matrix $\bC_\bma=\Ex{}{\bma \bma^\mathrm{H} }$. Furthermore, let $\boldsymbol{\ell}_1$ be the eigenvector associated with the largest eigenvalue $\lambda_1$ of the covariance matrix $\bC_\bma$. Then, the vector
\begin{align} \label{eq:solutiontooptimizationproblem2}
\hat\bmv = \boldsymbol{\ell}_1 +  \sign([\boldsymbol{\ell}_1]_i) \bme_i
\end{align}
is a solution of the following optimization problem: 
\begin{align} \label{eq:optimizationproblem2}
\hat{\bmv} \in \argmax_{\tilde\bmv\in\complexset^M }\, \Ex{}{|\bme_i^\mathrm{H} \bQ_{\tilde\bmv}\bma|^2}.
\end{align}
\end{lem}

We can now use \fref{lem:maxpowerisolation} to isolate the signal dimension with the highest receive power of antenna cluster $c$ on the first {pair of} AGCs/ADCs by setting $i=1$ and $\bma=\bmy_c$ with $\bC_\bma=\bC_{\bmy_c}=\Ex{}{\bmy_c\bmy_c^\mathrm{H}}$, which corresponds to solving
\begin{align}  \label{eq:specificoptimizationproblem2}
\hat{\bmv}^\mathrm{HR-MAX}_c \in \argmax_{\tilde\bmv_c\in\complexset^S }\, \Ex{}{|\bme_1^\mathrm{H} \bQ_{\tilde\bmv_c}\bmy_c|^2}.
\end{align}
According to \fref{lem:maxpowerisolation}, a vector that solves  \fref{eq:specificoptimizationproblem2} {is}
\begin{align}
\hat{\bmv}^\mathrm{HR-MAX}_c = \boldsymbol{\ell}_1 +  \sign([\boldsymbol{\ell}_1]_1) \bme_1,
\end{align}
where $\boldsymbol{\ell}_1$ is the eigenvector associated with the largest eigenvalue {$\lambda_1$} of the covariance matrix $\bC_{\bmy_c}$. Therefore, the HR matrix that focuses the signal dimension with the highest receive power on the first per-cluster {pair of} AGCs and ADCs is $\bQ_{\hat{\bmv}^{\mathrm{HR-MAX}}_c}$.
 

\section{Simulation Results} \label{sec:simulation_results}

We now demonstrate the efficacy of HDR MIMO in a mmWave massive MU-MIMO scenario where one UE has significantly stronger BS-side receive power than the other~UEs. 

\begin{figure*}[tp]
\centering
\subfigure[$\rho = 10$ dB]{
\includegraphics[width=0.315\linewidth]{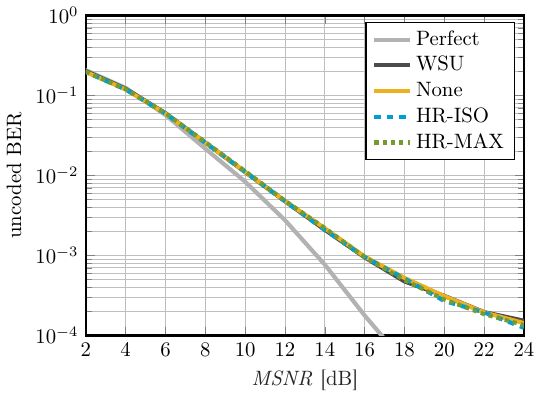} \label{fig:varying_rho_10}
}
\subfigure[$\rho = 20$ dB]{
\includegraphics[width=0.315\linewidth]{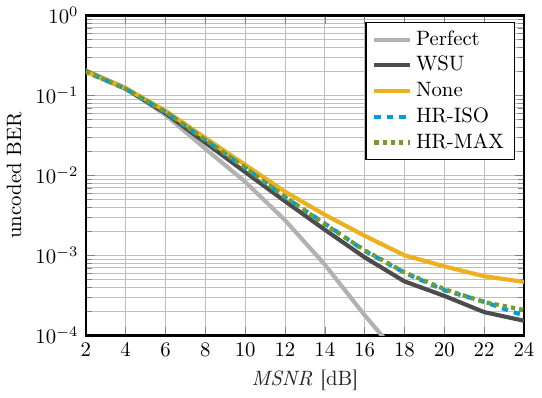}\label{fig:varying_rho_20}
}
\subfigure[$\rho = 30$ dB]{
\includegraphics[width=0.315\linewidth]{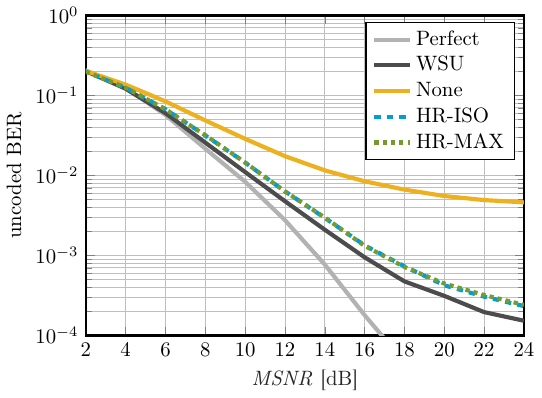}\label{fig:varying_rho_30}
}
\caption{Uncoded BER of {HDR MIMO and baselines} when varying the dynamic range $\rho$. In these results, $q = 3$ bit and $C = 32$ clusters.}
\label{fig:varying_rho}
\end{figure*}

\begin{figure*}[tp]
\centering
\subfigure[$q = 3$]{
\includegraphics[width=0.315\linewidth]{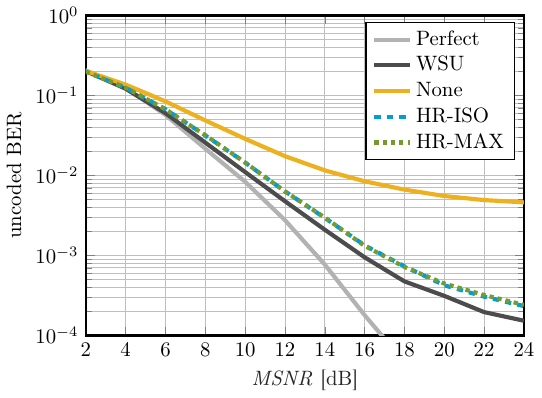}\label{fig:varying_q_3}
}
\subfigure[$q = 4$]{
\includegraphics[width=0.315\linewidth]{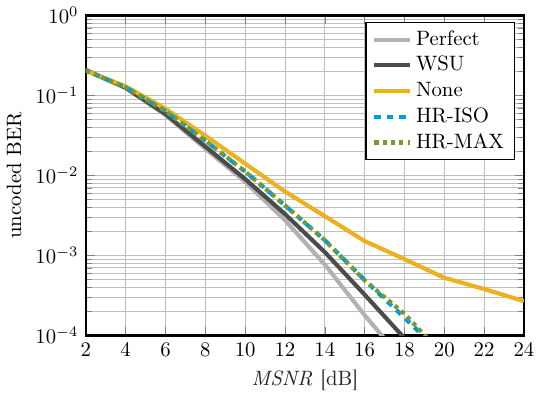}\label{fig:varying_q_4}
}
\subfigure[$q = 5$]{
\includegraphics[width=0.315\linewidth]{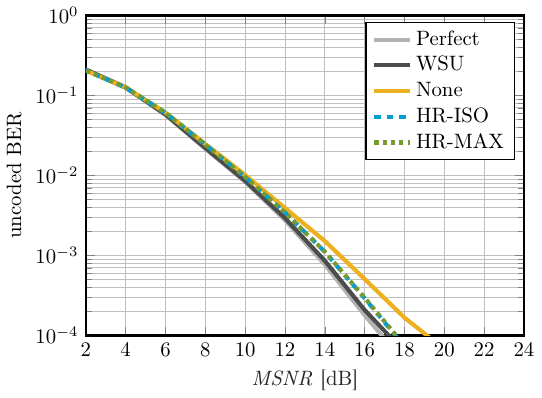}\label{fig:varying_q_5}
}
\caption{Uncoded BER of {HDR MIMO and baselines} when varying the number of quantization bits $q$. In these results, $\rho = 30$ dB and $C = 32$ clusters.}
\label{fig:varying_q}
\end{figure*}

\begin{figure*}[tp]
\centering
\subfigure[$C = 8$]{
\includegraphics[width=0.315\linewidth]{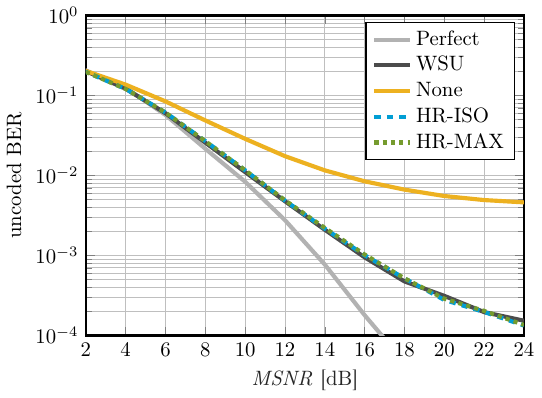}\label{fig:varying_C_8}
}
\subfigure[$C = 16$]{
\includegraphics[width=0.315\linewidth]{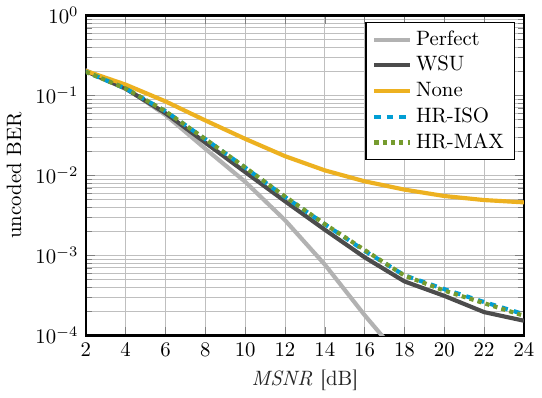}\label{fig:varying_C_16}
}
\subfigure[$C = 32$]{
\includegraphics[width=0.315\linewidth]{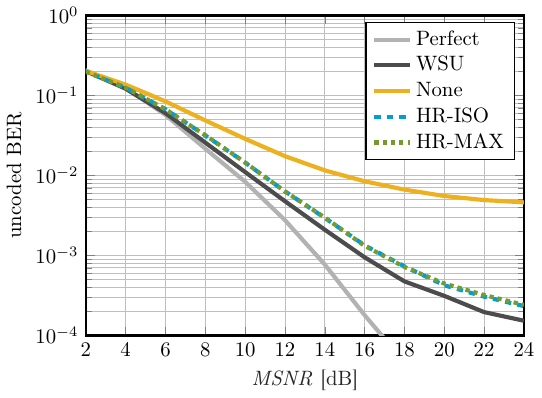}\label{fig:varying_C_32}
}
\caption{Uncoded BER of {HDR MIMO and baselines} when varying the number of antenna clusters $C$. In these results, $\rho = 30$ dB and $q = 3$ bit.}
\label{fig:varying_C}
\end{figure*}

\subsection{Simulation Details} \label{sec:simulated_scenario}

We consider an uplink scenario in which  $U=32$ single-antenna UEs transmit data to a BS with  $B=256$ antennas arranged as a uniform linear array (ULA) with half-wavelength antenna spacing. The BS is at a height of $25$\,m and the UEs at $1.5$m; both the BS and the UEs {are equipped with} omnidirectional antennas. The system operates at a carrier frequency of $60$\,GHz and a bandwidth of $40$\,MHz. The UEs transmit 16-QAM symbols. We limit the BS-side receive power between the weakest and the second strongest UE by $6$\,dB using the power control scheme in \cite[Sec.~II-B]{Song2021}, which determines the per-UE gains~$d_u$, $u=2,\dots,U$. In this scheme, strong UEs that exceed the dynamic range limit have to transmit with lower power while weak UEs transmit at their original power. We use mmWave channel vectors generated with Remcom's Wireless InSite~\cite{Remcom} for $30\,351$ UE positions in an area of $150\,\mathrm{m} \times 200\, \mathrm{m}$. For every channel realization, we select $U$ UE channel vectors uniformly at random. We define the median receive signal-to-noise ratio (MSNR) as follows:
\begin{align} \label{eq:snr_definition}
\textit{MSNR} =\frac{U \mathrm{median} \chav{\| \bmh_{u} \|^2,u=1,\ldots,U}}{B \No}.
\end{align}
The covariance matrix  $\bC_{\bmy}$ used in~\fref{eq:true_cy} for the AGC matrix~$\boldsymbol{\Omega}$  and in~\fref{eq:specificoptimizationproblem2} for solving HR-MAX is estimated as follows: $\bC_{\bmy}  \approx \frac{1}{K} \bY_T\bY^{\mathrm{H}}_T$. Here, the received matrix~$\bY_T$ is obtained during pilot-based training (cf. \fref{sec:system_model}), where we use orthogonal pilot sequences of length $K = U$ taken from a properly scaled Hadamard matrix.

\subsection{Baseline Algorithms}

We compare the proposed HDR MIMO methods {HR-ISO and HR-MAX} to three baselines. The first baseline is called ``Perfect'' and considers the same simulation setup as for HDR MIMO, but uses a receiver with infinite-resolution ADCs. Therefore, it does not require any spatial transform. The second baseline is called ``WSU'' (short for ``without strongest UE'') and considers a different simulation setup in which all UEs are power-controlled to a dynamic range of $6$\,dB and no spatial transform is used. The third baseline is called ``None'' and assumes the same simulation setup as for the HDR MIMO methods but without any spatial transform. Thus, we have $\bF = \bI_B$ for all baselines. 

\subsection{Simulation Results}
We evaluate the performance of HDR MIMO and the baseline algorithms in terms of uncoded {bit error rate (BER)}.We vary three parameters from our system: the dynamic range $\rho = \chav{10\,\text{dB}, 20\,\text{dB}, 30\,\text{dB}}$ from \fref{eq:rho_definition}, {the number of ADC quantization bits $q = \chav{3,4,5}$ from \fref{eq:uniform_midrise_quantizer}, and the number of antenna clusters $C = \chav{8,16,32}$ {from \fref{eq:y_tilde_c}}}. When varying one of these parameters, we fix all of the others to $\rho = 30$\,dB, {$q = 3$\,bit, and $C = 32$ {antenna} clusters}. 

In \fref{fig:varying_rho}, we show results varying the dynamic range~$\rho$. We observe that ``Perfect,'' which has infinite ADC resolution, outperforms all of the other presented methods which utilize 3-bit ADC resolution. As we increase the dynamic range $\rho$, the MSNR gap between {``WSU''} and the proposed HDR MIMO methods (HR-ISO and HR-MAX) increases. However, even with a $1000\times $ stronger BS-side receive power  between the weakest and strongest UEs, our proposed methods maintain an MSNR gap of only $1$\,dB at $0.1$\% BER between the HDR MIMO methods and {``WSU''}. Furthermore, the proposed HDR MIMO methods demonstrate significant performance gains compared to "None," especially for scenarios in which the BS-side receive power of the strongest UE far exceeds that of the other UEs (see \fref{fig:varying_rho_30}). Hence, the benefits of HDR MIMO are more pronounced with increasing BS-side receive power of the strongest UE.

In \fref{fig:varying_q}, we show results varying the number of quantization bits $q$. In \fref{fig:varying_q_3}, we observe a similar behavior as in \fref{fig:varying_rho}, where ``Perfect'' outperforms all of the other presented methods due to its infinite ADC resolution. As we increase the number of quantization bits $q$, the MSNR gap between ``Perfect,'' ``WSU,'' ``None,'' and the proposed HDR MIMO methods decreases, almost closing the MSNR gap in \fref{fig:varying_q_5}. This observation indicates that the benefits of HDR MIMO are less pronounced if the ADC resolution increases. Although with $5$-bit ADCs the HDR MIMO methods approach a nearly equivalent BER compared to the use of infinite-resolution ADCs, it also presents a smaller advantage when compared to ``None.'' Thus, HDR MIMO is suitable for systems with low-resolution ADCs.

In \fref{fig:varying_C}, we show results varying the number of antenna clusters $C$, and observe a similar behavior as the one from \fref{fig:varying_rho} and \fref{fig:varying_q_3}, where ``Perfect'' outperforms all of the other presented methods due to its infinite ADC resolution. As we increase the number of antenna clusters $C$, the MSNR gap between {``WSU''} and the HDR MIMO methods also increases. However, even when we have four times as many {antenna} clusters as in \fref{fig:varying_C_8}, we can still maintain a relatively small MSNR gap of only $1$\,dB at $0.1$\% BER between the HDR MIMO methods and {``WSU''} (\fref{fig:varying_C_32}). Furthermore, the HDR MIMO methods demonstrate significant performance gains compared to "None," especially in scenarios with fewer antenna clusters, as shown in \fref{fig:varying_C_8}. Hence, the benefit of HDR MIMO improves with fewer antenna clusters (which is equivalent to more antennas per cluster), even though it is disadvantageous for hardware implementation \cite{Enciso2020}.

Finally, we compare the BER performance between the two adaptive analog spatial transforms: HR-ISO and HR-MAX. In Figs.~\ref{fig:varying_rho}, \ref{fig:varying_q}, and \ref{fig:varying_C}, we find that the BER is nearly identical. In conclusion, both HDR MIMO methods are beneficial in systems featuring high dynamic range UEs and low-resolution ADCs. However, from a complexity perspective, HR-ISO is the preferable option as it does not require the calculation of the covariance matrix $\bC_{\bmy_c}$ and an eigenvalue decomposition.

\section{Conclusions}

We have proposed HDR MIMO, a novel approach that consists of an adaptive analog spatial transform which enables one to focus the energy of the {strongest UE} or the signal dimension with the highest receive power on a few ADCs in order to mitigate saturation and quantization artifacts caused by high dynamic range UEs in systems with low-resolution ADCs. Simulation results have shown that HDR MIMO greatly outperforms systems without spatial transforms. Furthermore, we have shown that (i) as the strongest UE {BS-side} receive power increases and (ii) as the quantization resolution decreases, the more pronounced the benefits of HDR MIMO become. Besides that, we have shown that, although transforming large antenna clusters is favorable, small analog spatial transforms that are more hardware friendly only  slightly deteriorate the effectiveness of HDR MIMO.


\appendices

\section{Proof of \fref{lem:HouseholderQR}}
\label{app:ISO}

The objective function of \fref{eq:optimizationproblem1} is bounded by 
\begin{align} \label{eq:cauchy_schwarz}
|\bme_i^\mathrm{H}\bQ_{\tilde\bmv} \bma|^{{2}} \overset{(a)}{\leq} \|\bme_i\|^{{2}} \|\bQ_{\tilde\bmv}\bma\|^{{2}} \overset{(b)}{=}  \|\bma\|^{{2}}.
\end{align}
Here,  $(a)$ follows from the Cauchy--Schwarz inequality and $(b)$ from the fact that $\bQ_{\tilde\bmv}$ is unitary. We now show that the {HR} matrix $\bQ_{\hat\bmv}$ with the vector $\hat\bmv$ from \fref{eq:solutionoptimizationproblem1} achieves the upper bound in \fref{eq:cauchy_schwarz} with equality. To this end, notice that the Cauchy--Schwarz inequality holds with equality if and only if $\bme_i$ and~$\bQ_{\hat\bmv}\bma$ are collinear, i.e., if $\alpha \bme_i = \bQ_{\hat\bmv}\bma$ holds for some $\alpha\in\complexset$. From \fref{eq:householder_matrix}, it follows that
\begin{align}  \label{eq:new1}
\alpha \bme_i = \bma -  2 \frac{\hat\bmv\hat\bmv^\mathrm{H}}{\|\hat\bmv\|^2}\bma.
\end{align} 
From \fref{eq:new1}, after some algebraic manipulations, we arrive at
\begin{align} \label{eq:new2}
\hat\bmv = \frac{\PC{\bma - \alpha\bme_i} \| \hat{\bmv}\|^2}{2 \hat{\bmv}^{\mathrm{H}}\bma},
\end{align}
where we assume that $\hat{\bmv}^{\mathrm{H}}\bma \neq 0$, i.e., $\hat{\bmv}$ and $\bma$ are not orthogonal. By inserting~$\hat\bmv$ from~\fref{eq:new2} into \fref{eq:new1}, the following must hold: 
\begin{align} \label{eq:new3}
\alpha \bme_i = \bma -  2 \frac{(\bma-\alpha\bme_i)(\bma-\alpha\bme_i)^\mathrm{H}}{\|\bma-\alpha\bme_i\|^2}\bma.
\end{align}
Since the factor {$\| \hat{\bmv}\|^2/(2 \hat{\bmv}^{\mathrm{H}}\bma)$} from \fref{eq:new2} cancels out, its specific choice does not matter (assuming that it is nonzero). Therefore, we assume {$\| \hat{\bmv}\|^2/(2 \hat{\bmv}^{\mathrm{H}}\bma) = 1$} in what follows.

In order to determine the optimal value of $\alpha$, we first assume that $\bma \neq \alpha \bme_i$. By rearranging terms in \fref{eq:new3}, we obtain the following necessary optimality condition
\begin{align}  \label{eq:interm_step}
\|\bma-\alpha\bme_i\|^2 =  2 (\bma-\alpha\bme_i)^\mathrm{H}\bma,
\end{align}
which can be simplified to 
\begin{align}   
 |\alpha|^2 &=  \|\bma\|^2- \alpha^* [\bma]_i +  \alpha [\bma]^*_i.
\end{align}
This equation has two solutions $\hat\alpha = \pm \|\bma\| \sign([\bma]_i)$. By inserting $\hat\alpha$ into~\fref{eq:new3}, both the left and right sides are equal, proving that the two solutions are indeed solutions to \fref{eq:new3}. For reasons given below, we pick the solution
\begin{align} \label{eq:alphaopt}
\hat\alpha = -\|\bma\| \sign([\bma]_i).
\end{align}

Let us now inspect the two excluded cases $\hat{\bmv}^{\mathrm{H}}\bma= 0$ and $\bma = \alpha \bme_i$. If $\hat{\bmv}^{\mathrm{H}}\bma= 0$, then the vectors $\hat{\bmv}$ and~$\bma$ are orthogonal. With \fref{eq:solutionoptimizationproblem1}, this implies that 
\begin{align}   
(\bma + \|\bma\| \sign([\bma]_i) \bme_i)^\mathrm{H} \bma = \|\bma\|^2+\|\bma\||[\bma]_i| = 0,
\end{align}
which is impossible as the vector $\bma$ was assumed to be nonzero;  this is the reason why we picked the solution in~\fref{eq:alphaopt}. If $\bma = \alpha \bme_i$, then according to \fref{eq:solutionoptimizationproblem1}, $\hat\bmv = 2\alpha \bme_i$. By plugging this vector into~\fref{eq:optimizationproblem1}, the bound in \fref{eq:cauchy_schwarz} is achieved with equality. 

In summary, by combining \fref{eq:new2} and \fref{eq:alphaopt}, we see that a vector $\hat\bmv$ that maximizes \fref{eq:optimizationproblem1} is given by \fref{eq:solutionoptimizationproblem1}. $\hfill\blacksquare$

\section{Proof of \fref{lem:maxpowerisolation}}
\label{app:MAX}

With the covariance matrix $\bC_\bma$, we can rewrite the optimization problem in \fref{eq:optimizationproblem2} as  
\begin{align} \label{eq:optimizationproblem4}
\hat{\bmv} \in \argmax_{\tilde\bmv\in\complexset^M }\,  \bme_i^\mathrm{H} \bQ_{\tilde\bmv}\bC_\bma \bQ_{\tilde\bmv} \bme_i.
\end{align}
We can now bound the objective of \fref{eq:optimizationproblem4} as follows:
\begin{align} \label{eq:upperbound}
\max_{\tilde\bmv\in\complexset^M }\,  \bme_i^\mathrm{H} \bQ_{\tilde\bmv}\bC_\bma \bQ_{\tilde\bmv} \bme_i \overset{(c)}{\leq} \max_{\bmz, \|\bmz\| = 1}\, \bmz^\mathrm{H}  \bC_\bma \bmz \overset{(d)}{=} \lambda_1.
\end{align}
Here, $(c)$ holds as $\bQ_{\tilde\bmv} \bme_i$ is a specific unit-norm vector and taking the maximum over arbitrary unit-norm vectors $\bmz\in\complexset^M$ bounds the objective. The equality $(d)$ follows from the definition of the spectral norm, which is equal to the largest eigenvalue~$\lambda_1$ of the covariance matrix $\bC_\bma$ of $\bma$~\cite[Ex.~5.6.6]{Horn1990}. We now show that the HR matrix~$\bQ_{\hat\bmv}$ with $\hat\bmv$ from~\fref{eq:solutiontooptimizationproblem2} achieves the upper bound in \fref{eq:upperbound} with equality. 

We start by rewriting \fref{eq:optimizationproblem4} as follows:
\begin{align}  \label{eq:newoptimizationproblem4}
\hat\bmv \in \argmax_{\tilde\bmv\in\complexset^M }\,  \sum_{m=1}^M \lambda_m \abs{\bme_i^\mathrm{H}  \bQ_{\tilde\bmv} \boldsymbol{\ell}_m}^2.
\end{align}
Here, the expression $\abs{\bme_i^\mathrm{H}  \bQ_{\tilde\bmv} \boldsymbol{\ell}_1}^2$ with $\tilde\bmv=\hat\bmv$ from~\fref{eq:solutiontooptimizationproblem2} is bounded~by
\begin{align} \label{eq:cauchy_schwarz_2}
|\bme_i^\mathrm{H}  \bQ_{\hat\bmv} \boldsymbol{\ell}_1|^2 \overset{(e)}{\leq} \|\bme_i\|^2 \|\bQ_{\hat\bmv} \boldsymbol{\ell}_1\|^2 \overset{(f)}{=}  1,
\end{align}
where $(e)$ follows from the Cauchy--Schwarz inequality and $(f)$ from the fact that $\bQ_{\hat\bmv}$ is unitary. From \fref{lem:HouseholderQR}, we know that using $\hat\bmv$ from \fref{eq:solutiontooptimizationproblem2} leads to equality  {in} $(e)$, i.e., $|\bme_i^\mathrm{H}  \bQ_{\hat\bmv} \boldsymbol{\ell}_1|^2 = 1$. This conclusion together with the objective from \fref{eq:newoptimizationproblem4} leads to
\begin{align}  \label{eq:lowerbound}
\lambda_1 \underbrace{|\bme_i^\mathrm{H}  \bQ_{\hat\bmv} \boldsymbol{\ell}_1|^2}_{1} + \sum_{m=2}^M \lambda_m \underbrace{\abs{\bme_i^\mathrm{H}  \bQ_{\hat\bmv} \boldsymbol{\ell}_m}^2}_{\geq 0} \geq \lambda_1,
\end{align}
since the eigenvalues $\lambda_m$, $m=2,\dots,M$ are nonnegative. By combining \fref{eq:upperbound} and \fref{eq:lowerbound}, we conclude that the objective function in \fref{eq:newoptimizationproblem4} is both lower and upper bounded by $\lambda_1$. Thus, with $\hat\bmv$ from \fref{eq:solutiontooptimizationproblem2}, we have that 
\begin{align} 
\sum_{m=1}^M \lambda_m \abs{\bme_i^\mathrm{H}  \bQ_{\hat\bmv} \boldsymbol{\ell}_m}^2 =  \lambda_1 \abs{\bme_i^\mathrm{H}  \bQ_{\hat\bmv} \boldsymbol{\ell}_1}^2 = \lambda_1,
\end{align}
which completes the proof. $\hfill\blacksquare$

\balance

\bibliographystyle{IEEEtran}
\bibliography{bib/VIPabbrv,bib/confs-jrnls,bib/publishers,bib/ASILOMAR_2023}


\end{document}